\documentclass[twocolumn,twoside,slac_two]{revtex4}
\usepackage{graphicx}
\usepackage{fancyhdr}
\begin{document}

\title{A Search for Millisecond Pulsars in \textit{Fermi} LAT-Detected Globular Clusters}

%

\author{Megan E. DeCesar}
\affiliation{Department of Astronomy, University of Maryland, College Park, MD 20740, USA}
\author{Scott M. Ransom}
\affiliation{National Radio Astronomy Observatory, Charlottesville, VA 22903, USA}

\author{Paul S. Ray}
\affiliation{U.S. Naval Research Laboratory, Washington, D.C. 20375, USA}

\begin{abstract}
We have searched for millisecond pulsars (MSPs) in two globular
clusters detected by the \textit{Fermi} Large Area Telescope. These
clusters contained no known MSPs prior to their detections in
$\gamma$-rays. The discovery of $\gamma$-ray emission from many MSPs
and the prevalence of MSPs in globular clusters points to a population
of MSPs as the likely source of the detected GeV emission, directing
our search for new cluster MSPs. We observed NGC 6652 and NGC 6388 at
2 GHz with the Green Bank Ultimate Pulsar Processing Instrument pulsar
backend at the Green Bank Telescope using the coherent dedispersion
mode. We have discovered one MSP in the $\gamma$-ray error circle of
NGC 6652. This pulsar is interesting because, while positionally
coincident with the GC, it has a much lower dispersion measure than
expected from the NE2001 galactic free electron density model. It is
unclear whether the MSP is a foreground pulsar or a cluster member,
and whether the pulsar, cluster, or both, is responsible for the
$\gamma$-ray emission. Timing the MSP will give the pulsar position
and a solid identification of the pulsar as a cluster member if it is
within a few core radii of the cluster center, as well as the
opportunity to search for $\gamma$-ray pulsations and determine the
origin of the GeV emission.

\end{abstract}

\maketitle

\thispagestyle{fancy}


\section{INTRODUCTION}

Globular clusters (GCs) contain orders of magnitude more low-mass
X-ray binaries (LMXBs) and recycled millisecond pulsars (MSPs) per
unit mass than the Galactic disk~\cite{Camilo_Rasio}. The high stellar
densities in GCs lead to high probabilities for stellar interactions
in these systems, through which binaries form and subsequently evolve.
GCs are the most likely hosts of exotic binary systems, like MSP-main
sequence binaries, highly eccentric binaries~\cite{Ransom2007}, and
MSP-black hole binaries, which would not form via normal stellar
evolution in the disk. Additionally, one can learn about cluster
dynamics and gas content, as well as MSP formation and evolution (due
to the high statistics), through the study of cluster pulsars.

The {\it Fermi Gamma-ray Space Telescope} Large Area Telescope
(LAT)~\cite{Atwood} has detected 18 previously known field MSPs at
energies $>100\,$MeV. Like normal pulsars~\cite{PSRCat}, all display a
spectrum consistent with an exponentially cutoff power law, with
cutoff energy $E_{\mathrm{c}} \sim\,$few GeV~\cite{Abdo_MSPs}. Radio
searches of unidentified \textit{Fermi} sources with pulsar-like
spectra have led to recent discoveries of 35 new MSPs
(e.g.~\cite{Ransom2011}).

While individual $\gamma$-ray MSPs would only be detectable in nearby
GCs or under special circumstances, a whole population of MSPs is
detectable at typical GC distances by the LAT~\cite{Venter2009}.
Fifteen GCs have been detected by the LAT thus
far~\cite{Kong2010}~\cite{LAT_GCs}~\cite{Tam2011}~\cite{2FGLCat}, and
those with the largest number of known MSPs, 47 Tucanae~\cite{47Tuc}
and Terzan 5, are the brightest of all clusters detected at GeV
energies. About half the GCs have spectra consistent with that of an
ensemble of MSPs~\cite{Venter2009}.

Nine of the LAT-detected GCs contain no known MSPs. As the LAT has
provided targets to search for MSPs in the field, we expect it is
pointing us to clusters that host MSPs but have remained undetected.
Many of these clusters are quite distant, of order 10 kpc away, and/or
are located farther south than is accessible by the Green Bank and
Arecibo telescopes, whose sensitivities are required to detect MSPs at
such distances. We chose to search for MSPs in two clusters, NGC 6388
and NGC 6652, with the Green Bank Ultimate Pulsar Processing
Instrument (GUPPI) pulsar backend at the Green Bank Telescope (GBT).
GUPPI is $\sim\,$twice as sensitive as SPIGOT, the previous pulsar
backend at the GBT, so may provide the extra sensitivity needed to
detect pulsars that were not detectable in past searches.

\section{TARGETED CLUSTERS}

The clusters we searched, NGC 6388 and NGC 6652, were chosen primarily
because of their high significance at GeV energies. 
They are very distant, at $\sim \,$11.6 and 9 kpc
respectively~\cite{LAT_GCs}. The dispersion measures (DM) of NGC 6388
and NGC 6652 are respectively $\mathrm{DM} \sim 345$ and
$196\,$cm$^{-3}\,$pc, according to the NE2001 model of galactic free
electrons~\cite{NE2001}. Typical predicted errors on the NE2001 DM
estimates are up to $\sim 50$\% for individual sources, though in some
cases the error can be larger (up to $\gtrsim100$\%). At such large
DMs, scattering is significant, and searching at a moderately high
frequency with coherent de-dispersion is the best recipe to reduce
scattering.

The GeV emission coincident with NGC 6388 has a hard power law
spectrum ($\Gamma \sim 1.1$) with a clear cutoff at $\sim 1.8
\,$GeV~\cite{LAT_GCs}, consistent with the combined spectrum from an
ensemble of $\gamma$-ray MSPs~\cite{Venter2009}. This cluster is the
most compact GC known~\cite{Harris}, and therefore has the highest
stellar encounter rate. Chandra observations show $\sim 60$ sources
within its half-mass radius, many of which are
LMXBs~\cite{Maxwell2007}. It is therefore expected that many MSPs
reside in this cluster, and only its distance has prevented their
detection.

NGC 6652 is firmly detected by the LAT, but no spectral cutoff was
measured by~\cite{LAT_GCs}. Its power law spectrum is hard, with a
photon index $\sim 1$. While this cluster is not very compact, it
contains 3 faint X-ray sources and a LMXB~\cite{Heinke}, suggesting
that MSPs likely exist in this cluster.

\section{OBSERVATIONS AND SEARCHES}

We observed both GCs at S-band (2 GHz) with the GUPPI backend on the
GBT between 2010 Oct 19 and 2011 May 6. The data have 800 MHz
bandwidth and 40.96$\, \mu$s time resolution. The observations were
performed in coherent search mode (e.g.~\cite{PSR_Handbook}), in which
the incoming data are dedispersed in real time at a pre-determined DM
to minimize pulse broadening. Each of the 2048 frequency channels
across the 800 MHz of bandwidth were coherently dedispersed at the
predicted DMs for the clusters. The channels were then combined in a
standard incoherent summation over a wide variety of search DMs. As
the coherent dedispersion search mode on GUPPI has only been available
for the past few years, these are some of the first coherent searches
that have been attempted.

These clusters are located at low declinations, and between
observatory scheduling constraints and the physical constraints of the
telescope, it was possible to observe them for only $\leq 3$ hr. NGC
6388 was observed 8 times, with individual observations lasting
between 1.5 and 2.5 hr. NGC 6652 was observed 6 times for 2--3 hr
each. The integration times limit our sensitivity to $\sim 20\,\mu$Jy
for NGC 6388 and $\sim 18\,\mu$Jy for NGC 6652, for the longest
observations.

All datasets, except one short observation of each cluster, have been
searched at the time of writing. The full data sets were processed
with PRESTO~\cite{Ransom_PhD}, which we used to dedisperse the time
series at $\sim\,$5000 DMs between 0 and 800 cm$^{-3}\,$pc and to
perform acceleration searches~\cite{Ransom2003} required to detect
most MSPs in binary orbits. We searched the full integrations with the
acceleration search parameter $z_{\mathrm{max}} = 200$. Additional
searches of shorter integration times and higher $z_{\mathrm{max}}$
values are underway.

\section{DISCOVERY OF A MSP COINCIDENT WITH NGC 6652}

\begin{figure*}
\includegraphics[scale=0.6,angle=-90]{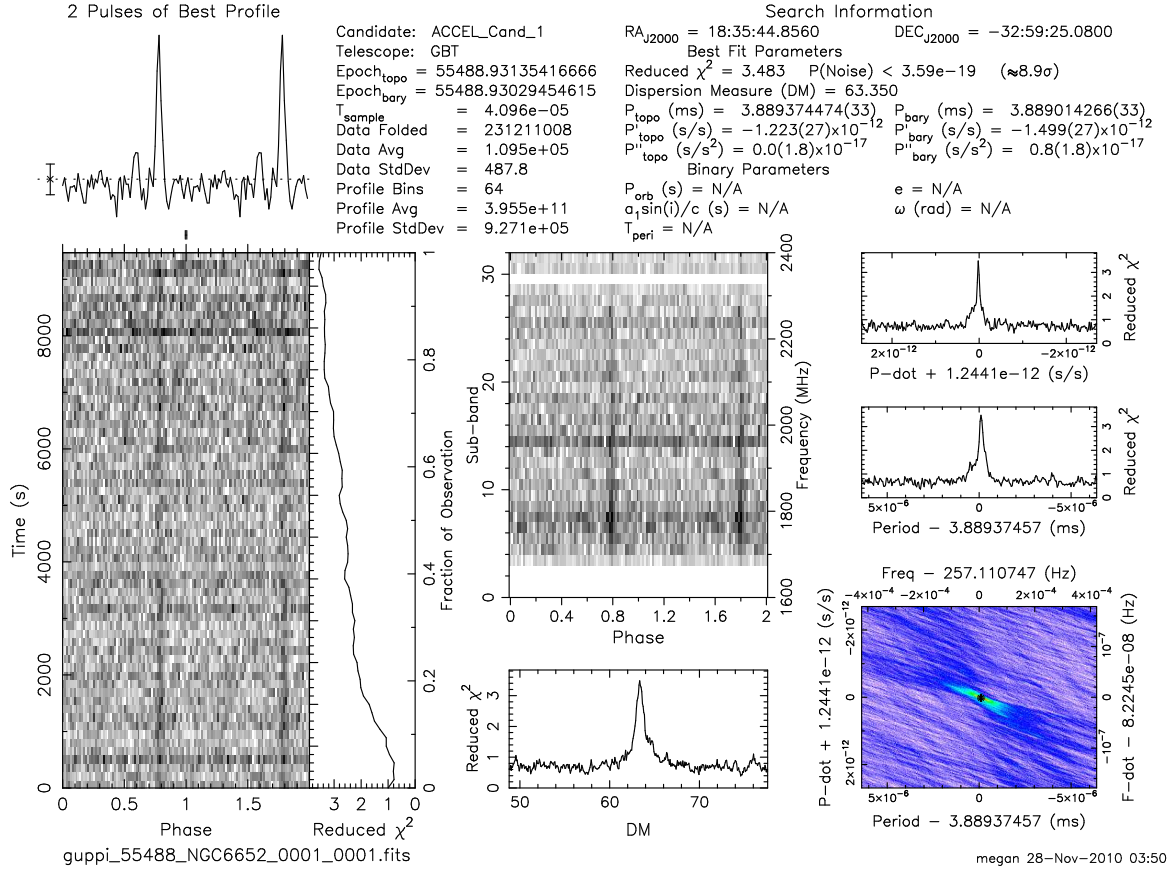}
\caption{The PRESTO discovery plot of PSR J1839$-$3259. Two cycles of
  the pulse profile integrated over time and frequency are shown in
  the upper left. The bottom left plot shows the persistence of the
  pulsar in time, while the middle plot shows the same across the
  bandwidth. The pulse signal-to-noise peaks in DM at
  63.35$\,$cm$^{-3}\,$pc, shown in the bottom middle plot. The peak in
  $\chi^2$ at the pulsar spin period and period derivative are shown
  in the plots on the right. The period derivative measured here is
  not the intrinsic $\dot{P}$ of the pulsar, but rather reflects the
  orbital acceleration of the pulsar during the observation.
  \label{ElizaPSR} }
\end{figure*}

Out of these searches has come the discovery of one new MSP thus far,
PSR J1839$-$3259. The discovery plot of this new pulsar is shown in
Figure~\ref{ElizaPSR}. The pulsar period is $\sim 3.89 \,$ms, and is clearly
accelerated; initial timing efforts suggest an orbital period of 1--2
days. Out of 6 observations, in only one is it not detected, perhaps
due to an eclipse.

This pulsar's DM is 63.35 cm$^{-3}\,$pc, much lower than the 196
cm$^{-3}\,$pc predicted in NE2001~\cite{NE2001}. This leads us to
question its association with NGC 6652. Judging from a {\it
SkyView}\footnote{http://skyview.gsfc.nasa.gov} image, the cluster's
angular diameter appears to be $\sim 2.2\,$arcmin. The GBT beam at
S-band is $\sim 6\,$arcmin. A chance coincidence between the cluster
field and pulsar location is therefore possible. The NE2001 model can
be very inaccurate, especially in the direction of the galactic center
where the cluster is located, so it is also possible that the expected
DM is very different from the true value. A discovery of a second
pulsar at the same DM would clinch the MSP as lying inside the
cluster, but searches at and around this DM for other pulsars have so
far been unsuccessful. Timing the MSP will reveal its true location,
as the phase-connected timing solution will include a very accurate
position that can be compared with the coordinates of the cluster
core. In addition, if it is in the cluster, there is a $\sim\,50$\%
chance that it will have a negative period derivative due to the
cluster potential.

The new radio MSP may also be a new $\gamma$-ray pulsar. The LAT 95\%
error radius of NGC 6652 is 7.2 arcmin, larger than the angular size
of the cluster and the S-band beam, so if the MSP is not in NGC 6652,
it could be a few arminutes from the cluster and still be coincident
with the cluster location at GeV energies. Upcoming observations will
allow us to obtain a timing solution, with which we will fold the LAT
photons to search for $\gamma$-ray pulsations. If found, we will look
in the off-peak phases to determine if any emission from NGC 6652
remains. If there is no remaining emission from NGC 6652, then the
pulsar is either in the foreground and the origin of the
$\gamma$-rays, or it is a cluster member but solely responsible for
the emission seen by the LAT, as in~\cite{Freire2011}.

\section{CONCLUSIONS}

The \textit{Fermi} LAT has led the way to the discoveries of many new
radio MSPs through pointed observations of unidentified $\gamma$-ray
sources. Following this lead, we searched for MSPs in two LAT-detected
globular clusters with the Green Bank Telescope, with the expectation
that each hosts a population of recycled pulsars.

We have discovered a new MSP coincident with the location of NGC 6652.
PSR J1839$-$3259 has a low DM of 63.35 cm$^{-3}\,$pc, much lower than
the NE2001 expectation of nearly 200 cm$^{-3}\,$pc, calling into
question its true association with NGC 6652. A timing campaign of this
pulsar will reveal whether or not it is a cluster member, and also if
it is a $\gamma$-ray pulsar or if the $\gamma$-rays are primarily from
the GC.

No cluster MSPs have conclusively been discovered from these
observations thus far. While we have found many pulsar candidates near
the expected DM values for the two GCs we searched, none repeat from
one observation to the next, which is essential for confirming a
pulsar detection. Scintillation is not an issue at such large
distances and does not explain the lack of detections. The scattering
timescale for NGC 6652 is $\tau_{\mathrm{scatt}} \sim 0.7\mu$s, and
$\tau_{\mathrm{scatt}} \sim 3\mu$s for NGC 6388~\cite{NE2001}. These
numbers change very little when the DM is doubled to account for
potential errors in the NE2001 model, so scattering probably does not
play a role in our findings. It is most likely that the distances to
the clusters, combined with the relatively short integration times due
to their low declinations, are limiting our sensitivity to the MSPs
that undoubtedly reside in these systems. Further observations and
reprocessing of the data with higher acceleration searches may still
yield detections of MSPs from these clusters. Additionally, several
more clusters with no known MSPs have recently been detected by the
LAT~\cite{Tam2011}. We will continue our search for cluster MSPs with
the GBT in these newly $\gamma$-ray-detected globular clusters.

\bigskip 
\begin{acknowledgments}

The National Radio Astronomy Observatory is a facility of the National
Science Foundation operated under cooperative agreement by Associated
Universities, Inc.

The \textit{Fermi} LAT Collaboration acknowledges support from a
number of agencies and institutes for both development and the
operation of the LAT as well as scientific data analysis. These
include NASA and DOE in the United States, CEA/Irfu and IN2P3/CNRS in
France, ASI and INFN in Italy, MEXT, KEK, and JAXA in Japan, and the
K.~A.~Wallenberg Foundation, the Swedish Research Council and the
National Space Board in Sweden. Additional support from INAF in Italy
and CNES in France for science analysis during the operations phase is
also gratefully acknowledged.

The authors thank F. Camilo at Columbia University for the use of his
computer cluster.

MED acknowledges Eliza Jolene Bailey, whose birth coincided with the
first observation of NGC 6652 that led to the detection of PSR
J1839$-$3259.


\end{acknowledgments}

\bigskip 

\begin{thebibliography}{99} 



\bibitem{47Tuc}
Abdo et al. 2009, Science 325, 845.

\bibitem{Abdo_MSPs}
Abdo et al. 2009, Science 325, 848.

\bibitem{LAT_GCs}
Abdo et al. 2010, A\&A 524, 75.

\bibitem{PSRCat}
Abdo et al. 2010, ApJS 187, 460.

\bibitem{2FGLCat}
Abdo et al. 2011 (arXiv:1108.1435)

\bibitem{Atwood}
Atwood et al. 2009, ApJ 697, 1071.

\bibitem{Camilo_Rasio}
Camilo \& Rasio 2005, ASP Conf. Series 328, 147.

\bibitem{NE2001}
Cordes \& Lazio 2002 (arXiv:astro-ph/0207156)

\bibitem{Freire2011}
Freire et al., submitted.

\bibitem{Harris}
Harris 1996, AJ 112, 1487.

\bibitem{Heinke}
Heinke 2004, PhDT.

\bibitem{Kong2010}
Kong et al. 2010, ApJ 723, 1219.

\bibitem{PSR_Handbook}
Kramer \& Lorimer 2005, Handbook of Pulsar Astronomy.

\bibitem{Maxwell2007}
Maxwell et al. 2007, AAS 211, 03.16

\bibitem{Ransom_PhD}
Ransom 2001, PhDT.

\bibitem{Ransom2003}
Ransom 2003, ApJ 589, 911.

\bibitem{Ransom2007}
Ransom 2007, IAU Symposium 246, 291.

\bibitem{Ransom2011}
Ransom et al. 2011, ApJ 727, 16.

\bibitem{Tam2011}
Tam et al. 2011, ApJ 729, 90.

\bibitem{Venter2009}
Venter et al. 2009, ApJ 696, 52.




\end{thebibliography}

\end{document}